\documentstyle [12pt,float,epsf,epsfig,amssymb,amsmath]{article}

\begin{document}
\def\l{\lambda}
\def\m{\mu}
\def\a{\alpha}
\def\b{\beta}
\def\g{\gamma}
\def\d{\delta}
\def\e{\epsilon}
\def\o{\omega}
\def\O{\Omega}
\def\v{\varphi}
\def\t{\theta}
\def\r{\rho}
\def\bs{$\blacksquare$}
\def\bp{\begin{proposition}}
\def\ep{\end{proposition}}
\def\bt{\begin{th}}
\def\et{\end{th}}
\def\be{\begin{equation}}
\def\ee{\end{equation}}
\def\bl{\begin{lemma}}
\def\el{\end{lemma}}
\def\bc{\begin{corollary}}
\def\ec{\end{corollary}}
\def\pr{\noindent{\bf Proof: }}
\def\note{\noindent{\bf Note. }}
\def\bd{\begin{definition}}
\def\ed{\end{definition}}
\def\C{{\mathbb C}}
\def\P{{\mathbb P}}
\def\Z{{\mathbb Z}}
\def\d{{\rm d}}
\def\deg{{\rm deg\,}}
\def\deg{{\rm deg\,}}
\def\arg{{\rm arg\,}}
\def\min{{\rm min\,}}
\def\max{{\rm max\,}}

\newtheorem{th}{Theorem}[section]
\newtheorem{lemma}{Lemma}[section]
\newtheorem{definition}{Definition}[section]
\newtheorem{corollary}{Corollary}[section]
\newtheorem{proposition}{Proposition}[section]

\begin{titlepage}

\begin{center}

\topskip 5mm

{\LARGE{\bf {Smooth Parametrizations in

\vskip 3mm

Dynamics, Analysis, Diophantine and

\vskip 3mm

Computational Geometry}}}

\vskip 8mm

{\large {\bf Y. Yomdin$^{*}$}}

\vspace{4 mm}
\end{center}

{$^{*}$ Department of Mathematics, The Weizmann Institute of
Science, Rehovot 76100, Israel}

\vspace{2 mm}

{e-mail: yosef.yomdin@weizmann.ac.il}

\vspace{1 mm}

\vspace{1 mm}
\begin{center}

{ \bf Abstract}
\end{center}
{\small{Smooth parametrization consists in a subdivision of the mathematical objects under consideration into simple pieces,
and then parametric representation of each piece, while keeping control of high order derivatives. The main goal of the present
paper is to provide a short overview of some results and open problems on smooth parametrization and its applications in several
apparently rather separated domains: Smooth Dynamics, Diophantine Geometry, Approximation Theory, and Computational Geometry.

The structure of the results, open problems, and conjectures in each of these domains shows in many cases a remarkable similarity,
which we try to stress. Sometimes this similarity can be easily explained, sometimes the reasons remain somewhat obscure, and it
motivates some natural questions discussed in the paper. We present also some new results, stressing interconnection between
various types and various applications of smooth parametrization.

}}

\vspace{2 mm}
\begin{center}
------------------------------------------------
\vspace{2 mm}
\end{center}
This research was supported by the ISF, Grant No. 779/13, and by the Yeda-Sela Foundation.

\end{titlepage}
\newpage

\section{Introduction}
\setcounter{equation}{0}

``Parametrization'' is a change of variables which simplifies understanding of a mathematical structure under investigation.
The most important example in the realm of Algebraic and Analytic Geometry is provided by Resolution of Singularities, in its
various versions (\cite{Hir,Bie.Mil} and references therein). In many problems of Dynamics, Analysis, Diophantine and
Computational Geometry it is important to keep control of high order derivatives while performing a change of variables.
Parametrizations of this type, which we call ``smooth parametrizations'', is the main topic of the present paper.

\smallskip

The main example is provided by the ``$C^k$-parametrization theorem'' (\cite{Gro1,Pil.Wil,Yom1,Yom2}). This can be considered as
a high order quantitative version of the well known result on the existence of a triangulation of semi-algebraic sets, with the
number of simplices bounded in terms of the degree (see \cite{Hir}). In a $C^k$-version we just require in addition, that each
simplex be an image of the standard one, under the parametrization mapping $\psi$, with all the derivatives of $\psi$ up to
the order $k$ uniformly bounded. We shall consider below also ``Mild parametrizations'' (\cite{Mas,Jon.Tho,Jon.Mil.Tho,Pil4,Tho1,Tho2}),
where all the derivatives up to infinity are controlled, and ``Analytic parametrizations'' (\cite{Yom3,Yom7}) where the norm of an
analytic extension into the complex domain is controlled.

\smallskip

Today smooth parametrizations are ``traditionally'' applied in Dynamics and in Diophantine Geometry (Section \ref{Dyn.Appl} and
\ref{Dioph.Geo} below). Important applications in Computational Geometry have been proposed
(\cite{Hav.Yom,Xu.Mou.Duv.Gal1,Xu.Mou.Duv.Gal2,Xu.Mou.Duv.Gal3} and Section \ref{Comp.Geom} below). Also in Approximation Theory
(especially in study of Polynomial Inequalities on algebraic sets) importance of parametrizations was well recognized
(\cite{Bar.Ple1,Bar.Ple2,Bar.Ple3,Yom9} and Section \ref{Remez} below).

\smallskip

One can trace a certain similarity in the structure of the results, open problems, and conjectures in each of these domains of
applications of smooth parametrization. The main goal of the present paper is to provide a short overview of the current results
and open problems in smooth parametrization and its different applications, stressing their similarity (sometimes striking). We do
not try to present a comprehensive review of either of the domains, and the references are kept to a minimum (still exceeding an
average in similar texts, because of an attempt to cover several fields).

Some new results are also presented, in Sections \ref{non.rigid}, \ref{Comp.Geom}, and \ref{Remez}.

\smallskip

The author would like to thank D. Burguet, G. Comte, O. Friedland, Y. Ishii, G. Jones, G. Liao, P. Milman, B. Mourrain, J. Pila,
R. Pierzchava, M. Thomas, A. Wilkie for useful discussions, and for explaining him some topics presented below. Special thanks 
belong to RIMS Institute, Kyoto, and to the organizers of the conference there in July 2013 on semi-algebraic techniques in Dynamics, 
which inspired a good part of this paper.

\section{What sets can we parametrize?}
\setcounter{equation}{0}

In this paper we discuss parametrization of semi-algebraic sets, sets definable in $o$-minimal structures, and functions with a bounded
number of zeroes.

\subsection{Semi-algebraic sets}

Semi-algebraic sets in ${\mathbb R}^n$ are defined by a finite number of real polynomial equations and inequalities, and set-theoretic
operations. See, for example, \cite{Ben.Ris,Yom.Com}. Assume we are given a semi-algebraic set $A\subset {\mathbb R}^n$. A diagram
$D(A)$ of the set $A$, is the collection of the ``discrete'' data of $A$, i.e. the degrees and the number of the equations and
inequalities, and the set-theoretic formula defining $A$. So $D(A)$ does not depend on specific values of the coefficients of the
polynomials involved.

\subsection{$o$-minimal structures}\label{omin}

We do not give here a formal definition of $o$-minimal structures (see \cite{Van,Wil} and references therein). This notion was developed
in Model Theory, and turned out to be very useful in Analysis, Differential Equations and Diophantine Geometry. Informally, $o$-minimal
structure $\cal S$ over ${\mathbb R}$ consists of collections $S_n$ of subsets in ${\mathbb R}^n, \ n\geq 1,$ closed under set-theoretic
operations, products and projections. It is required that each $S \in S_1$ is a finite union of intervals, closed or open. For
$X \subset {\mathbb R}^n$ we say that $X$ is definable in $\cal S$ if $X \in S_n$. Usually semi-algebraic sets are assumed to belong to
$\cal S$.

\smallskip

One important example of an $o$-minimal structure is provided by globally subanalytic sets. This structure is denoted ${\mathbb R}_{an}.$
Real semi-analytic sets are obtained from real analytic ones in the same way as semi-algebraic sets are constructed from
algebraic ones. But this class is not closed under proper projections, so the images under such projections are added. See \cite{Bie.Mil}.

\smallskip

Another example is the structure ${\mathbb R}_{exp},$ where exponential functions are added to polynomials. Here $S_n$ is the collection of
subsets in ${\mathbb R}^n$ of the form $X=\pi(f^{-1}(0))$, with $\pi:{\mathbb R}^m \to {\mathbb R}^n$ the projection on the first $n$
coordinates, and $f$ an exponential polynomial in $x_1,\ldots,x_m$, i.e. polynomial in $x_1,\ldots,x_m,e^{x_1},\ldots,e^{x_1}$. Note
that ${\mathbb R}_{exp}$ contains sets such as $\{(x,e^{-{1\over x}}), \ x>0\},$ which are not subanalytic at the origin. A deep fact is
that ${\mathbb R}_{exp}$ is an $o$-minimal structure (\cite{Wil}).

\smallskip

The finiteness assumption guarantees that a kind of ``real Bezout theorem'' is valid in any $o$-minimal structure $\cal S$: if an
intersection of definable sets consists of isolated points, then their number is finite.

\section{Main types of smooth parametrization}
\setcounter{equation}{0}

To simplify the presentation we give here all the main definitions for real semi-algebraic sets. Later we extend this setting to
$o$-minimal structures, and (in Section \ref{non.rigid} below) to much less rigid assumptions of ``zero counting''.

\bd\label{rep}
A parametrization of $A$ is a subdivision of $A$ into semi-algebraic pieces $A_j$ together with algebraic mappings (``charts'')
$\psi_j: I^{n_j}\to A_j$, where $I^{n_j}$ is the cube $[-1,1]^{n_j}$ in ${\mathbb R}^{n_j}$. We assume additionally that $\psi_j$ are
onto and homeomorphic on the interiors of $I^{n_j}$ and $A_j$.
\ed
In some applications it is enough to assume that the images of the charts cover $A$, and not to require a subdivision.

\smallskip

Now, three main types of smooth parametrization - the $C^k$-one, the $C^\infty$ mild parametrization, and the analytic parametrization
- differ between them in the requirements imposed on the charts.

\subsection{$C^k$-parametrization}

\bd\label{Ckchart}
A $C^k$-mapping $\psi: I^{n}\to A$ is called a $C^k$-chart, if its $C^k$-norm $\Vert\psi(x)-\psi(0)\Vert_{C^k}$ is bounded by 1. A
parametrization of $A$ is called a $C^k$-one, if all the mappings $\psi_j: I^{n_j}\to A_j$ are $C^k$-charts.
\ed

The following result (in a weaker form) was originally obtained in \cite{Yom1,Yom2}. M. Gromov obtained a full version in \cite{Gro1}.
Today detailed proofs are available in \cite{Bur1,Pil.Wil}.

\bt\label{mainck}
For any natural $k$ and for any compact semi-algebraic set $A$ inside the cube $I^n$ in ${\mathbb R}^n$, there exists a
$C^k$-parametrization of $A$, with the number of $C^k$-charts, depending only on $k$ and on the diagram $D(A)$ of $A$.
\et
One feature of this result is that the number of charts depends only on the differentiability order $k$ and on the ``combinatorial''
characteristics of $A$: the degrees of the polynomials involved and the set-theoretic formula. In the applications in Dynamics, for which
this result was initially intended, this requirement cannot be relaxed. Any dependence of the number of charts on the specific values of
the coefficients of the arising polynomials would completely destroy the proof.

\smallskip

In some other applications the ``uniformity'' requirement can be relaxed. However, in the presentation below we always stress the
parameters controlling the number of charts in each specific version of smooth parametrization.

\smallskip

Theorem \ref{mainck} was extended to $o$-minimal structures in \cite{Pil.Wil}:

\bt\label{Pil.Wil} (\cite{Pil.Wil}, Theorem 2.3) Let $\cal S$ be an $o$-minimal structure. For each $k \in {\mathbb N}$ and each bounded
$X$ definable in $\cal S$ there exists a $C^k$- parametrization of $X$.
\et
This basic result shows that much less than semi-algebraicity is required for a validity of $C^k$- parametrization. However, in
general we do not have a notion of a ``degree'' (or of a ``diagram'', or of any other kind of a ``combinatorial complexity'' of a set) in
$o$-minimal structures. So the problem of {\it uniform bounds} on the size of zero-dimensional definable sets can be studied only in
specific cases (compare, however, Section 2 of \cite{Pil.Wil}).

\smallskip

A related result on $o$-minimal stratifications was obtained in \cite{Fis}.

\smallskip

Very recently, a version of $C^k$ parametrization theorem was obtained in \cite{Clu.Com.Loe} for $p$-adic definable sets, and more broadly,
in a non-archimedean, definable context. In particular, piecewise approximation by Taylor polynomials was extended in \cite{Clu.Com.Loe} to
this setting. This result was applied in \cite{Clu.Com.Loe} to bounding the number of rational points of a given height on the transcendental
part of $p$-adic subanalytic sets.

\subsection{Mild parametrization}

Mild parametrization has been introduced in \cite{Pil5} and further studied in \cite{Pil6,But,Tho1,Tho2,Jon.Tho,Jon.Mil.Tho} and other
publications. The following definitions are slightly modified versions of Definitions 2.1, 2.3, and 2.4 from \cite{Pil6}:

\bd\label{Mild.chart}
A $C^\infty$-mapping $\psi: I^{n}\to {\mathbb R}^n$ is called an  $(A,C)$-mild chart, if for each multi-index $\alpha \in {\mathbb N}^n$
we have

\be\label{Mild.eq}
\max_{x\in I^n}\Vert \partial^\alpha \psi(x)\Vert \leq \alpha!(A|\alpha|^C)^{|\alpha|}.
\ee
\ed

\bd\label{rep.mild}
An $(A,C,N)$-mild parametrization of a set $V\subset I^m$ is a collection of $N$ $(A,C)$-mild charts, whose images are contained in $V$
and cover it.
\ed
Mild parametrizations form an important tool in study of density of rational and algebraic points on analytic varieties. In Section
\ref{Dioph.Geo} we present shortly some recent results of in this direction.

\smallskip

In \cite{Pil6} it is conjectured that mild parametrization exists for sets and functions definable in algebraic-exponential $o$-minimal
structure. Some special cases of this conjecture are proved in \cite{Pil6}. Some other cases are presented in the papers cited above.

\smallskip

However, in \cite{Tho1} an example was presented of an $o$-minimal structure, not allowing for mild parametrization.

\smallskip

It was shown in \cite{Pil6} that mild parametrization exists for globally subanalytic sets - this follows from ``resolution of singularities''
for such sets, obtained in \cite{Bie.Mil}. However, the uniformity of the number of charts in mild parametrization of semi-algebraic sets
seems to be an open (and important) question. Analytic parametrization, considered in the next section, is automatically mild, but it is known
(see \cite{Yom7} and Section \ref{Repp.2.alg} below) that {\it the required number of analytic charts depends on the coefficients of the
polynomials involved}. In particular, for $A$ being the part of the hyperbola $xy=\e^2$ inside the box $I^2$, we need at least
$C\log ({1\over \e})$ analytic charts. How many mild charts do we need here?

\subsection{Analytic parametrization}\label{Analytic.Par}

\bd\label{achart}
An ``analytic $K$-chart'' (or, shortly, an ``a-K-chart") is a
real analytic mapping $\psi : I^n \to {\mathbb R}^m,$ such that $\psi$ is extendible, as a complex analytic
mapping, to the concentric complex polydisk $\Delta^n_3 = \{(z_1,\dots, z_n) \in {\mathbb C}^n, \ \vert z_i \vert \leq 3, \ i=1,
\dots, n \},$ and the norm of \ $\psi(z)-\psi(0)$ is bounded in $\Delta^n_3$ by $K$. An ``analytic $1$-chart" will be called 
shortly an a-chart.

A parametrization of $A\subset {\mathbb R}^n$ is called an analytic one, if all its charts $\psi_j$ are a-charts.
\ed

\bt\label{res.sing}
For any globally subanalytic set $A$ there exists an analytic parametrization of $A$ with a finite number of charts. In particular,
this is true for bounded semi-algebraic sets $A$ in ${\mathbb R}^n$.
\et
\pr
This follows from the uniformization theorem of \cite{Bie.Mil}: $A$ can be analytically parametrized by a finite number of
compact nonsingular analytic varieties. The last can be covered by finite number of analytic coordinate charts, which, in turn, can
be subdivided into a-charts. $\square$

\smallskip

{\it An essentially new feature of analytic parametrization is that already for semi-algebraic sets $A$ we cannot expect the number
of a-charts to be bounded in terms of the diagram $D(A)$}

\smallskip

So, for analytic parametrization (as well as for $C^k$ and mild ones) there are three questions, natural from the point of view of
applications: {\it How far can we go beyond semi-algebraic sets? How many charts do we need? What level of uniformity for the number
$N$ of charts can we expect?}

\smallskip

The proof of Theorem \ref{res.sing} stress the importance in this context of understanding of the complexity of the uniformization of
globally subanalytic sets. There are two recent papers \cite{Bie.Mil.Gri.Wlo,Gri.Mil} where some results in this direction are presented. 

\smallskip

Another approach, for semi-algebraic sets $A$ in ${\mathbb R}^2,$ was suggested in \cite{Yom7}: to get back a uniform bound on $N$ depending
only on $D(A)$, we exclude from covering a few boxes in $A$ of size $\delta >0$, this $\delta$ being an additional parameter of the
problem. As its is clear from the proof in two-dimensional case, provided in Section \ref{An2} below, the excluded boxes have to cover
``complex singularities'' of $A$.

\bd\label{anal.rep.def}
Let $\delta > 0$. An analytic $\delta$-parametrization of a set  $A \subset I^n$ is an analytic parametrization of the set 
$A\setminus \cup_{j=1}^N W_j,$ where $W_j$ are open boxes of size at most $2\delta$.
\ed

\bt\label{anal.rep.th} (\cite{Yom8}, Theorem 3.1) Let $A\subset I^2$ be a compact semi-algebraic set, and let $\delta > 0$. 
There exist an analytic $\delta$-parametrization of of $A$ with the following properties:

\smallskip

\noindent 1. The number $N$ of the removed boxes is bounded by a constant $C_1$.

\noindent 2. The number of charts is bounded by $C_2 \log({1\over \delta})$.

\noindent 3. Each chart is algebraic, and its degree is bounded by $C_3$.

\smallskip

\noindent The constants $C_1,C_2,C_3$ depend only on the diagram $D(A)$.
\et
It was shown in \cite{Yom1,Yom2,Yom3} that omitting from the covering of $A$ these $\delta$ boxes is allowed in Dynamical applications,
as far as  the bounds are logarithmic in $\delta$. Indeed, we can put $\delta$ to be ``exponentially small''. Presumably, this trick
can work also in other applications.

\medskip

Notice that analytic parametrization is automatically mild. Indeed, the bounds on the derivatives of the a-chart, given by the
Cauchy formula are the following: for $\psi : I^n \to {\mathbb R}^m$ an a-$K$-chart we have for each partial derivative of the multi-order
$\alpha=(\alpha_1,\dots,\alpha_n)$ the bound $\vert {d^{\alpha}\psi \over {dz^{\alpha}}}\vert \leq {{CK}\over {2^{\vert \alpha \vert}}},$
which is stronger than that of (\ref{Mild.eq}).

\smallskip

On the other hand, since analytic parametrization is also mild, the example of \cite{Tho1} shows that there are $o$-minimal structures
without analytic parametrization. A natural question here is whether a $\delta$-parametrization exists in Thomas' example? Notice also,
that the existence of analytic parametrization in a certain {\it real structure} is closely related to counting {\it complex zeroes} (compare
Section \ref{non.rigid} below).

\medskip

Extending the result of Theorem \ref{anal.rep.th}, (and Theorem \ref{anal.rep.fns.th} below) to higher dimensions seems to be a difficult
problem. One of the reasons is that the geometry of complex singularities of $A$ in dimensions higher than 2 may be fairly complicated. Notice
that we cannot expect in general analytic parametrization outside of a finite number of $\delta$-boxes, with only $\log {1\over \delta}$ of
a-charts:
consider hyperbolic cylinder $xy=\e^2$ in $I^3\subset {\mathbb R}^3$. For $1 >> \delta >> \e$, whatever number of $\delta$-boxes we delete,
a part of the $z$-axis will remain uncovered, and near this part we need an order of $\log {1\over \e} >> \log {1\over \delta}$ a-charts.
Presumably, an approach of \cite{Yom2}, which settles in higher dimensions a similar difficulty, can be used also in the analytic case.

\section{About proofs in $C^k$ and analytic cases}
\setcounter{equation}{0}

We present here some steps in the proof of the $C^k$ and analytic parametrization theorems in two-dimensional case, i.e. for semi-algebraic
sets in ${\mathbb R}^2$. Our goal is to illustrate the similarities and the differences between these two cases, and to stress the role of
``zeros counting''. Notice that the one-dimensional case of the parametrization result is immediate: a closed semi-algebraic set
in $[-1,1]\subset {\mathbb R}$ is a finite union of closed intervals; each of these intervals can be linearly parametrized by the unit interval.

\subsection{$C^k$-parametrization in dimension 2}\label{Repp.2}

Let $A$ be a compact semi-algebraic set inside the cube $I^2$.
After a simple subdivision with certain vertical, horizontal, and diagonal 
straight lines, we may assume that $A_j$
has either the form $\{(x,y)\in I^2 =[-1,1]^2, \ g_1(x) \leq y
\leq g_2(x) \}$, where $0\leq g_1(x) < g_2(x)\leq 1$ are two
regular algebraic functions on $I=[-1,1]$, satisfying $\vert g'_i
\vert \leq 1, \ i=1,2$, or a symmetric with respect to the
coordinates $x,y$ form. Some $g_i$ may be constant.
If, in addition, we had all the derivatives of $g_1$ and $g_2$ up
to the order $k$ bounded by $1$, we could parametrize $A$
using the following $C^k$-chart $\psi: I^{2}\to A$:
\begin{equation}
\psi(t_1,t_2)= (t_1, t_2g_2(t_1)+(1-t_2)g_1(t_1)).
\end{equation}
Therefore, it is enough to prove that each regular algebraic function $g(x)$ on $I$ can be parametrized by a partition of the interval
and by subsequent changes of the independent variable in such a way that all its derivatives up to $k$ become small. We prove this fact
in the next section.

\subsection{$C^k$-parametrization of algebraic functions}\label{Repp.2.alg}

\bd\label{rep.fn}
A $C^k$-parametrization of an algebraic function $g(x)$ on
$I=[-1,1]$ is a partition of $I$ into subsegments $\Delta_j$
together with the collection of \ $C^k$-charts $\psi_j:I\to
\Delta_j$ such that $g \circ \psi_j: I \to {\mathbb R}$ are also
$C^k$-charts.
\ed
We shall prove the following result:
\bp\label{prop.rep}
Let $g(x)$ be a regular algebraic function of degree $d$ on $I$ satisfying
$0 \leq g(x) \leq 1$ and $\vert g'(x)\vert \leq 1, \ x \in I$. Then there
exists a $C^k$-parametrization of $g$ with the number of the
partition intervals bounded through $d$.
\ep
\pr
We mark in $I$ all the
zeroes of all the derivatives of $g$ up to order $k+1$, subdivide
$I$ by all the marked points and linearly reparametrize each of
the subdivision intervals by $I$. So we may assume that all the
derivatives of $g$ up to order $k$ preserve their sign and are
monotone on $I$.

\medskip

We continue by induction on the number of the consecutive
derivatives of $g$ which are already ``small''. By assumptions of
the proposition, the first derivative $g'$ already satisfies
$\vert g'(x)\vert \leq 1, \ x \in I$. So let us assume that all
the consecutive derivatives $g^{(i)}$ of $g$ up to order $l-1, \ 1
\leq l-1 < k$ satisfy $\vert g^{(i)}(x)\vert \leq 1, \ x \in I$,
and consider the next derivative $g^{(l)}(x)$. By the
construction, $g^{(l)}(x)$ does not change sign and is monotone on
$I$. We can assume, for example, that it is positive and
monotonously decreasing.

\bl\label{lemma1}
The $l$-th derivative of $g$, \ $g^{(l)}(x)$ satisfies on $[0,1]$ the
inequality $g^{(l)}(x) \leq {1\over x}$.
\el
\pr
Otherwise, if for a certain $x_0\in [0,1], \ \ g^{(l)}(x_0)
> {1\over x_0}$, then, by monotonicity, we have $g^{(l)}(x) >
{1\over x_0}$ for each $x \leq x_0$. Integrating the last
inequality on the interval $[0,x_0]$ we get $
g^{(l-1)}(x_0)-g^{(l-1)}(0) > 1$, which contradicts the
induction assumptions.

\medskip

Now we perform a nonlinear change of variables which finally
"kills" the $l$-th derivative of $g$: put $h(t)=t^2, \ t\in [0,1]$
and consider the composition $\hat g(t)=g(h(t)).$

\bl\label{lemma2}
All the consecutive derivatives $\hat g^{(i)}$ of $\hat g$ up to order $l$
satisfy $\vert \hat g^{(i)}(x)\vert \leq C, \ x \in [0,1]$, with
the constant $C$ depending only on $l$.\el \pr Write an expression
for the $i$-th derivative of the composition $g(h(t))$, using the
chain rule. We see that for $i < l$ all the terms in the resulting
expression are uniformly bounded, and hence this derivative does
not exceed $C(l)$. For the $l$-th derivative of this composition
we have
\begin{equation}\label{lemma2.eq}
{d^l(g(h(t))\over {dt^l}}=g^{(l)}(h(t))\cdot
(2t)^l+R(t),
\end{equation}
where $R(t)$ contains only the derivatives of $g$
up to the order $l-1$, and hence $R(t)$ is uniformly bounded. For
the first term in (\ref{lemma2.eq}), by Lemma \ref{lemma1} we have $g^{(l)}(h(t))\leq
{1\over h(t)} ={1\over t^2}$. Since $l \geq 2$, the first term in
(\ref{lemma2.eq}) does not exceed $2^l$. This completes the proof of Lemma
\ref{lemma2}.

\medskip

To complete the proof of Proposition \ref{prop.rep} we notice that the change
of variables applied has a fixed degree $2$. Hence, after each its
application we get a new algebraic function of the degree at most
twice larger than of the original one. Now we repeat, if
necessary, a subdivision of the interval $I$, in order to remove
possible sign changes of the derivatives, and apply the next
induction step. After $k$ steps we "kill" all the derivatives of
$g$ up to order $k$, while the total number of the subdivision
intervals $\Delta_j$ remains bounded by the degree of $g$. By
construction, the degree of the parametrizing mappings is
bounded by $2^k$. $\square$

\subsection{An example: parametrization of $H=\{xy+\e^2=0\}$}

Consider the component of $H_{\e}$ over the negative $x$-axis. First of
all, we subdivide this component into two symmetric pieces by the
point $(-\e, \e)$. Consider the piece with $-1 \leq x\leq -\e.$
The second piece is parametrized in a symmetric way. So we have
to $C^2$-parametrize the algebraic function $g(x)={-\e^2 \over x}$ on the interval $[-1,-\e]$. We see immediately, that
{\it all the derivatives of $g(x)$ are positive over the interval $[-1,-\e]$.} So the step of subdivision in the proof 
of Proposition \ref{prop.rep} above is not necessary.

\medskip

Hence, the nonlinear change of variables we have to apply, takes
the form $x= h(t)= -t^2-\e, \ t\in [\sqrt {1-\e},0]$. The first derivative
of $g(x)$ is bounded by one, and the same is true for the
first derivative $\hat g'(t)=g'(h(t))h'(t)$ of $\hat g(t)=g(h(t))$. For the second derivative we have 
$\hat g''(t)=g''(h(t))(h'(t))^2+g'(h(t))h''(t)$, and according to the
above computation, $$\vert \hat g''(t) \vert \leq 2 + 2g'(h(t)) \leq 4.$$ So it is enough to subdivide the interval 
$[\sqrt {1-\e},0]$ into two equal pieces and to rescale them linearly by $I$ in order to reduce the bound to $1$.

Explicitly, we have $\tilde g_{\e}(t)={\e^2\over {t^2+\e}}$, and a simple direct calculation confirms the above estimate.

\subsection{Analytic $\delta$-parametrization in dimension 2}\label{An2}

The proof goes basically in the same lines as the proof of Theorem \ref{mainck} given above: it is reduced to a parametrization
of an algebraic function $g(x)$ of one variable.

\bd\label{anal.rep.fns}
Let $\delta > 0$ and a real algebraic function $g(x)$ on $I=[-1,1]$ be given. An analytic $\delta$-parametrization of $g(x)$
consists of the following objects:

\medskip

1. A finite number of open subintervals $U_i$ of $I$, $i=1,...,N$, with the length of each $U_i$ at most $2\delta$.

\medskip

2. A partition of $I\setminus \cup^N_{i=1}U_i$ into subsegments $\Delta_j, \ j=1,\dots,M,$ together with the collection of a-charts
$\psi_j:I\to \Delta_j$ such that $g \circ \psi_j: I \to {\mathbb R}$ are also a-charts.
\ed
We have the following result:

\bt\label{anal.rep.fns.th} (\cite{Yom7}, Theorem 3.4)
There are constants $C_1(d)$ and $C_2(d)$ such that for each real algebraic function $g(x)$ of degree $d$ on $I$ satisfying
$0 \leq g(x) \leq 1, \ x \in I$, and for each $\delta > 0$, there is an analytic $\delta$-parametrization of $g$ with the number
$N$ of the removed intervals bounded by $C_1(d),$ and the number $M$ of the partition intervals bounded by
$C_2(d)\log {1\over \delta}.$ All the a-charts in this parametrization are affine.
\et
\pr
Consider a complete analytic continuation $\hat g(z)$ of the algebraic function $g(x)$ from $I$ to the complex plane $\mathbb C$.
In general, $\hat g(z)$ is a multivalued analytic function (with at most $d$ branches) outside of its singular set
$\Sigma=\{z_1,\dots, z_m\} \subset {\mathbb C}$, $m\leq d(d-1)$.

\medskip

Now, in contrast to the $C^k$ case, in analytic parametrization we have to avoid complex singularities of $\hat g$: if an a-chart
comes ``too close'' to a singularity, it must, in fact, cover it, which contradicts the definition of a-charts. Consequently, the
size of the images of a-charts decreases as the distance to the nearest singularity. This implies a necessity of the number of the
partition elements of order $C'(d)\log {1\over \delta}$.

\smallskip

More accurately, let $\delta > 0$ and let the points $z_1,\dots, z_m \in {\mathbb C}$ be given. Denote by $x_1,\dots,x_m$ the projections
of the points $z_1,\dots, z_m$ to the real line, and denote by $U^i_{\delta}, \ i=1,\dots,m$ the open $2\delta$-intervals centered at
$x_i$.

\bl\label{real.partition}
For any $\delta > 0$ and the points $z_1,\dots, z_m \in {\mathbb C}$ the complement $I\setminus \cup^m_{i=1}U^i_{\delta}$ can be
covered by not more that $2(m+1)log_2({1 \over {\delta}})$ intervals $\Delta_j$ with the following property: the distance of
the central point $c_j$ of $\Delta_j$ to each of $z_1,\dots, z_m$ is not smaller than three times the length $\vert \Delta_j \vert$
of the interval $\Delta_j$.
\el
\pr
The complement $I\setminus \cup^m_{i=1}U^i_{\delta}$ consists of at most $m+1$ intervals $J_r, \ r=1,\dots,s \leq m+1$. For each
interval $J_r$, in order to subdivide it into the required subintervals $\Delta_j$, we proceed as follows: we take the interval of
the length $\delta \over 4$ from the left of $J_r$, next to it we take the interval of the length $\delta \over 2$, next to it the
interval of the length $\delta$, then $2\delta$, and so on, until we cover the central point of $J_r$. Then we repeat the same
construction from the right. Clearly, we need $m\log {1 \over {\delta}}$ subintervals. The details are given in \cite{Yom7}.
$\square$

\smallskip

For each interval $\Delta_j$ consider now the open disk $D^j$ of radius $3\vert \Delta_j \vert$ centered at the central point $c_j$
of $\Delta_j$. The function $\hat g$ is regular on $D^j$ and by assumptions it is bounded by $1$ on $\Delta_j$. To prove Theorem
\ref{anal.rep.fns.th} it remains to apply the following ``Bernstein inequality for algebraic functions'' (\cite{Roy.Yom,Yom7}):

\bp\label{Bern.Ineq}
Let $\hat g$ be an algebraic function of degree $d$, univalued and regular in the disk $D_{3R}$ and bounded in absolute value by
$1$ on the real interval $[-R,R]$. Then $\hat g$ is bounded in absolute value by $C(d)$ on the disk $D_{2R}$.
\ep
Thus for an affine mapping $\psi: D_1\to D^j$ the composition $g\circ \psi$ is a $C(d)$-chart. A further subdivision provides the
required number of $a$-charts. This completes the proof of Theorem \ref{anal.rep.fns.th}. $\square$

\subsection{Smooth parametrization and zero counting}\label{non.rigid}

While in this paper we mostly consider smooth parametrization of {\it semi-algebraic sets}, and, to some extent, of sets definable in
a certain $o$-minimal structure, it is pretty clear from the proofs in Sections \ref{Repp.2} and \ref{An2} that in fact much less is
required (at least, in dimensions one and two): uniform bounds on zeroes of the considered functions and their derivatives. This fact 
is used, in particular, in some recent bounds on the density of rational points on analytic curves (compare \cite{Pila7,But}). For $C^{k+1}$ 
functions on $[-1,1]$ we get

\bt\label{Smooth.Ck.th}
Let $f(x)$ be a $C^{k+1}$ function defined for $x\in [-1,1]$ satisfying the condition $\max_{x\in [-1,1]]} |f'(x)|\leq 1$. Assume that the number
of zeroes of each of $f,f',\ldots,f^{(k+1)}$ in $[-1,1]$ does not exceed $N$. Then there exists a $C^k$-parametrization of $f$ with the number
of charts, depending only on $N$.
\et
\pr
It is identical to the proof of Proposition \ref{prop.rep}. We notice that the algebraicity assumption on $g$ in this proposition was used
only in order to provide an upper bound on the number of zeroes of $g$ and its derivatives. $\square$

\medskip

However, for analytic parametrization of real analytic functions $f$ we need a certain bound on ${\it complex}$ zeroes of $f$. The following
definition (\cite{Yom9}) slightly extends the classical notion of ``valency'' of $f$ (compare \cite{Hay} and references therein):

\bd\label{SPvalDef}
A function $f$ regular in a domain $\O\subset {\mathbb C}$ is called $p$-valent in $\O$ if for any $c \in {\mathbb C}$ the number of solutions
of the equation $f(x)=c$ in $\O$ does not exceed $p$. The function $f$ is called $(s,p)$-valent in $\O$ if for any polynomial $P(x)$ of degree
at most $s$ the number of solutions of the equation $f(x)=P(x)$ in $\O$ does not exceed $p$.
\ed
For $s=0$ we obtain the usual $p$-valent functions. Easy examples (see \cite{Yom9}) show that an $(s,p)$-valent function may be not
$(s+1,p)$-valent.

\smallskip

Algebraic functions and solutions of linear ODE's with polynomial coefficients are $(s,p)$-valent for each $s$ and an appropriate $p=p(s),$
away from their singularities. But the class of $(s,p)$-valent functions is much wider and much ``less rigid'' than those. In particular, it
was shown in \cite{Bat.Yom} that functions whose Taylor coefficients satisfy a linear Poincar\'e-type recurrence relation, are $(s,p)$-valent,
for each $s$ and $p=p(s)$.

\smallskip

We shall consider a class $S(n,p_1,p_2)$ of meromorphic functions $f$ in a complex disk $D_3,$ real on the real line, with at most $n$ poles
$z_1,\ldots,z_n \in D_3$, which are $p_1$-valent, and also $(p_1,p_2)$-valent in $D_3\setminus \{z_1,\ldots,z_n\}.$

\bt\label{anal.rep.s.p.th}
There exist constants $C_1(n,p_1,p_2)$ and $C_2(n,p_1,p_2)$ such that for each function $f \in S(n,p_1,p_2)$ satisfying
$0 \leq f(x) \leq 1, \ x \in I$, and for each $\delta > 0$, there exists an analytic $\delta$-parametrization of $f$ with the number $N$ of the
removed intervals bounded by $C_1(n,p_1,p_2)$ and the number $M$ of the partition intervals bounded by $C_2(n,p_1,p_2)log_2({1\over \delta}).$
All the a-charts in this parametrization are affine.
\et
\pr
It is identical to the proof of Theorem \ref{anal.rep.fns.th}. We notice that the algebraicity assumption on $g$ in this theorem was used only
in order to justify application of the Bernstein inequality (Proposition \ref{Bern.Ineq} above). However, in \cite{Yom9} Bernstein inequality
was extended to the functions in $S(n,p_1,p_2)$. $\square$

\section{Applications of smooth parametrization}
\setcounter{equation}{0}

In this section we outline shortly some by now ``traditional'' applications of smooth parametrization in Dynamics and in Diophantine
Geometry. We suggest also a new possible field of applications: Polynomial Approximation, in its more theoretical aspects (Remez-type
inequalities), and in more applied ones (Computational Geometry). Let us just mention some other important applications of smooth
parametrization, in particular, in \cite{Bou.Gol.Sch,Bur2,Gro2}, in directions which we do not discuss in this paper.

\subsection{Smooth Dynamics}\label{Dyn.Appl}

Let $f:X\to X$ be a continuous mapping of a compact metric space
$X$. For $n=0,1,\dots$ define a metric $d(f,n)$ on $X$ as
$$d(f,n)(x,y)=\max_{i=0,1,\dots,n}d(f^{\circ i}(x),f^{\circ
i}(y)),$$ where $d$ is the original metric on $X$ and $f^{\circ
i}=f\circ f\circ \dots \circ f$ denotes the $i$-th iteration of
$f$.

\medskip

For $\e>0$ let $M(f,n,\e)$ denote the minimal number of $\e$-balls
in $d(f,n)$-metric, covering $X$. Notice that the $\e$-ball
$B^n_{\e}$ centered at $x\in X$ in $d(f,n)$-metric consists of all
$y\in X$ such that $d(f^{\circ i}(x),f^{\circ i}(y))\leq \e, \
i=0,1,\dots,n$. So, the orbits of $x$ and $y$ till $n$
must remain in a distance at most $\e$.

\medskip

We expect an exponential in $n$ growth of the covering number
$M(f,n,\e)$, so we define the $(n,\e)$-entropy $h(f,n,\e)$ of $f$
as the rate of this growth: $$h(f,n,\e)=n^{-1}log_2 M(f,n,\e).$$
Finally, the topological entropy $h(f)$ is defined as
$$h(f)=\lim_{\e \to 0} \overline{\lim_{n \to \infty}}h(f,n,\e).$$
Computation of the topological entropy $h(f)$ and investigation of
its behavior is usually difficult because of the complicated
geometry of the $\e$-balls $B^n_{\e}$ and of the irregular
character of the two limit processes involved.

However, the $(n,\e)$-entropy $h(f,n,\e)$ of $f$ should be
considered as a ``computable" quantity, although the complexity of
the required computations grows exponentially in $n$. Thus, it is
important to estimate the ``remainder term'' $r(f,n,\e)=h(f)-h(f,n,\e)$.

\smallskip

Let us assume now that $f:M\to M$ is a smooth mapping of a compact smooth manifold $M$. One can show that in this case $r(f,n,\e)$
is bounded by the ``local volume growth'' $LV(f,\e)$ (or, better, a ``local complexity growth'' $LC(f,\e)$, in the spirit of
Gromov's definition in \cite{Gro1}). The first is the maximal exponential rate of the growth under iterations of $f$ of the
volume of the part of submanifolds inside the $\e$-balls $B^n_{\e}$ in $d(f,n)$-metric. Local complexity growth is defined in a
similar way.

\smallskip

Our goal is to show that regularity assumptions on $f$ imply upper bound on $LC(f,\e)$: for $f$ in $C^k$ we expect
$\lim_{\e\rightarrow 0} LC(f,\e)\leq {m\over k}\log L(f).$ Here $L(f)$ is the largest ``Lyapunov exponent'' of $f$. (One cannot get better
bounds: easy examples show that already for a linear $f$ there are $m$-dimensional $C^k$-submanifolds whose local volume growth under
iterations of $f$ is ${m\over k}\log L(f)$).

One of the main difficulties in the analysis of the local complexity growth in iterations of a non-linear $f$ is that the geometric
complexity of the $\e$-balls $B^n_{\e}$ in $d(f,n)$-metric grows exponentially with $n$. Another difficulty, which prevents a
straightforward application of the $C^k$- regularity, is that the $C^k$-norm $||f^{\circ n}_{C^k}||$ of the iterations of $f$ grows with
$n$ as $L(f)^{nk}$. In contrast, for a linear $f$ this growth is only of the order $L(f)^n$, and in this case one can easily show, that
the local volume growth under iterations of $f$ for $m$-dimensional $C^k$ - submanifolds is at most ${m\over k}\log L(f)$.

\smallskip

Applying $C^k$-parametrization on each iteration simplifies the geometry, preserving control of higher derivatives, thus settling
the first difficulty. Then working with ``blocks'' $f^{\circ q}$ instead of $f$, and rescaling with respect to $\e\to 0$ ``kills''
higher derivatives, resolving the second difficulty. So we obtain the following result:

\bt\label{entck}(\cite{Yom1})
Let $f:M\to M$ be a $C^k$ mapping. Then $\lim_{\e\rightarrow 0} LC(f,\e)\leq {m\over k}\log L(f).$ In particular, for $f\in C^\infty$
we have $\lim_{\e\rightarrow 0} LC(f,\e)=0.$
\et
This result has many important consequences for smooth dynamics (see \cite{Bur3,Bur.Lia.Yan,DeT.Vig,Gue,Lia,Lia.Via.Yan,McM,Mon,New1,New2,Yom1}
and references therein). In particular, it confirms, for $C^\infty$-mappings, the so called ``Entropy conjecture'', which asks for a lower
bound for $h(f)$ through the spectral radius of $f_*$ acting on the homology of $M$. It also implies upper semicontinuity of the topological
entropy $h(f)$ and existence of invariant measures with maximal metric entropy (\cite{New2}). In smooth and holomorphic
dynamics Theorem \ref{entck} is mostly used as a ``black box''. However, in some applications, especially, in rational dynamics, entering the 
proof and its modification is required (compare \cite{DeT.Vig}). In particular, this may concern specifying the kind of smooth parametrization 
used.

\smallskip

The result of Theorem \ref{entck} leaves many important questions open. Arguably, the most prominent is the validity of the
Entropy conjecture for $f$ in the class $C^{1+\e}$ - see \cite{Lia.Via.Yan} and references therein. There are important problems in smooth
Dynamics, which require a sharpening of the bounds on the complexity growth, obtained in Theorem \ref{entck}. In particular, this
concerns the explicit bound of $C(f,\e)$ for $C^\infty$ or analytic $f$, closely related explicit bounds of the semi-continuity modulus
of the topological entropy in $C^\infty$ and analytic families, the problem of estimating the topological entropy in finite accuracy
computations (compare \cite{Ish.San,Mil,New.Ber.Gro.Mak}), as well as the problem of bounding entropy for rational maps with singularities
(\cite{DeT.Vig,Gue}).

\smallskip

Replacing $C^k$-parametrization with analytic one answers some of these questions. One of the main results of \cite{Yom3}, obtained on the
base of a low-dimensional analytic parametrization, is the following:

\bt\label{anal.ent}
Let $f:M\to M$ be a real analytic diffeomorphism of a compact real analytic surface $M$.
Then $$r(f,n,\e)\leq C(f,\e)\leq C \ {{\log \ \log {1\over \e}}\over {\log {1\over \e}}}.$$
\et
Similar bounds can be obtained for the semi-continuity modulus of the topological entropy.

\medskip

On the base of these low-dimensional results we've proposed in \cite{Yom3} the following conjecture:

\smallskip

\noindent{\bf Conjecture 1} For real analytic mappings $f:M\to M$ of compact real
analytic manifolds $M$ of any dimension always $C(f,\e) \leq C \ {{\log \ \log {1\over \e}}\over {\log {1\over \e}}}$.

\smallskip

Very recently Conjecture 1 was proved in \cite{Bur.Lia.Yan}, with only $C^k$-parametrization techniques, combined with an accurate
control of the growth with $k$ of the $k$-th order derivatives. Moreover, this result has been obtained also for $f$ quasi-analytic.
Earlier a uniform upper bound for $C(f,\e)$ for analytic maps was obtained in \cite{Lia}.

\smallskip

There is a strong similarity between these results and certain results in bounding rational points: see the next section.

\smallskip

Presumably, extending analytic parametrization to higher dimensions will provide another proof of Conjecture 1. We expect that it will be
useful in many other dynamical problems, in particular, in Computational Dynamics, where we want to numerically calculate iterations of
nonlinear maps with a prescribed accuracy, and with a minimal computational effort. Dynamical partitions are well known to be an efficient
tool in such computations (compare ``Taylor models'', as used in \cite{New.Ber.Gro.Mak,Wit.Ber.Gro.Mak.New}). One can hope that $C^k$ and
analytic parametrization can be used in the same lines, improving ``resolution'' of the computations.

\subsection{Diophantine Geometry}\label{Dioph.Geo}

In the origin of many exciting recent developments in Diophantine Geometry, in particular, those involving smooth parametrization, was
a remarkable paper by E. Bombieri and J. Pila \cite{Bom.Pil}. In this paper elementary, but highly delicate techniques in Calculus and
Algebraic Geometry were used to give upper bounds for the number of integer points on the graphs of functions $y=f(x)$ under various
smoothness and convexity hypotheses.

We state here two of the basic ``preliminary'' results of \cite{Bom.Pil} on the images of smooth mappings, and one of arithmetic
conclusions, in order to illustrate the way in which smooth parametrization enters the bounds on density of rational points. Then we
mention very shortly some other results in this line, mainly in order to stress an apparent similarity of the results and open problems
in density of rational points, and in other fields which use smooth parametrization.

\smallskip

Let $n,m \in {\mathbb N}$ be fixed. Assume that $\phi=(\phi_1,\ldots,\phi_m):I^n \to {\mathbb R}^m$ is a $C^\infty$ mapping, and for
each $k\in {\mathbb N}$ let $M_k(\psi)$ denote the norm $||\phi||_{C^k}$. Let $z^1,\ldots,z^m \in I^n$ be any $m$ points. The first basic
lemma of \cite{Bom.Pil} bounds from above the generalized Vandermonde determinant $\Delta$ of $\phi$, where
$\Delta= \det (\phi_i(z^j)), \ i,j=1,\ldots,m.$ (In this and two further statements our presentation follows \cite{Pil3,Mar}).

\smallskip

Let $D_s(l)=(^{s+l}_{\ \ s})$ (respectively, $L_s(l)=(^{s+l-1}_{\ \ s-1})$) be the dimension of the space of polynomials (respectively, of
homogeneous polynomials) of degree $l$ in $s$ variables $x=(x_1,\ldots,x_s)\in {\mathbb R}^s$. We shall use the degree of differentiability
$k=k(n,m)$ of $\psi$, where $k$ is uniquely defined by the requirement $D_n(k)\leq m <D_n(k+1)$.

\smallskip

Finally, define $e(n,m)$ by $e=\sum_{l=0}^k L_n(l)\cdot l +(k+1)[m-D_n(k)].$

\bl\label{BP1}
Let $B^n_r$ ba a ball of radius $r<1$ in ${\mathbb R}^n$. Then for each $z^1,\ldots,z^m \in I^n \cap B^n_r$ and for
$\Delta= \det (\phi_i(z^j)), \ i,j=1,\ldots,m,$ we have

\be\label{BP1.eq}
|\Delta|\leq m! [D_n(k)M_k(\psi)]^m r^e
\ee
\el
\pr
Apply Taylor approximation of order $k$ of $\phi$ and use linear dependence between the monomials in $n$ variables $x=(x_1,\ldots,x_n)$
of degree at most $k$. $\square$

\smallskip

For $\O$ a bounded subset in ${\mathbb R}^m$ and $t\in {\mathbb R}$ let $t\O({\mathbb Z})$ be the set of integer points in the
$t$-dilation $t\O$ of $\O$. We denote by $\O(t,{\mathbb Z})$ the set of points $y\in \O$ such that $ty \in t\O({\mathbb Z})$.

\bp\label{BP2}
Let $\phi=(\phi_1,\ldots,\phi_m):I^n \to {\mathbb R}^m$ be as above, and let $\O\subset {\mathbb R}^m$ be the image of $\psi$. Then for each
$d \in {\mathbb N}$ there are constants $C(\psi,d)$ and $\e(\psi,d)$, such that $\e(\psi,d)\to 0$ as $d\to \infty$, with the following
property: for each $t$ the set $\O(t,{\mathbb Z})$ is contained in the union of at most $C(\psi,d)t^{\e(\psi,d)}$
algebraic hypersurfaces of degree less than or equal to $d$.
\ep
\pr
For each multi-index $\a=(\a_1,\ldots,\a_m)$ with $|\a|\leq d$ consider the monomial $\eta_\a(y)= y^\a=y_1^{\a_1}\cdot \ldots \cdot y_m^{\a_m}$
on ${\mathbb R}^m$. We denote by $\tau=D_m(d)$ the number of these monomials, and fix a certain ordering $\a^i, \ i=1,\ldots,\tau,$ of the
multi-indices $\a$. The mapping $V_d=(\eta_{\a^1},\ldots,\eta_{\a^\tau}):{\mathbb R}^m \to {\mathbb R}^\tau$ is called the Veronese mapping of
degree $d$.

\smallskip

Let $W=\{w^1,\ldots,w^s\} \in \O$ be a finite subset. We form the Vandermonde matrix $A=(\eta_{\a_i}(w_j)), \ i=1,\ldots,\tau, \ j=1,\ldots, s.$
It is well known that $W$ is contained in an algebraic hypersurface of degree less than or equal to $d$ if and only if the rank of $A$ is less
than $\tau$ (see, for example, \cite{Bom.Pil}, Lemma 1, or \cite{Bru.Yom}, Proposition 2.2).

\smallskip

Now the idea is to apply Lemma \ref{BP1} to the smooth mapping $\psi=V_d\circ \phi: I^n \to {\mathbb R}^\tau$. So we put $\tilde k=k(n,\tau)$,
and $\tilde \e = e(n,\tau)$, as it was defined in Lemma \ref{BP1}. Notice that in fact $\tilde k$ and $\tilde \e$ depend only on $n,m,d$.

\smallskip

Let us fix $\tilde r$ such that $\tau! [D_n(\tilde k)M_{\tilde k}(\psi)]^\tau \tilde r^e < t^{-\kappa}$, where
$\kappa=\sum_{l=0}^\tau L_n(l)\cdot l.$ Thus $\tilde r = C_1 t^{-{\kappa \over e}}.$ We cover $I^n$ by balls $B_j$ of radius
$\tilde r$. We need $C_2 ({1\over {\tilde r}})^n = C_3 t^{{{\kappa n} \over e}}= C(\psi,d)t^{\e(\psi,d)}$ such balls, with $C(\psi,d)=C_3,$
$e(\psi,d)={{{\kappa n} \over e}}$.

Let for some $j$ a finite set $W=\{w^1,\ldots,w^s\}$ be contained in $\O \cap \phi(B_j).$ Denote by $z^1,\ldots,z^s \in I^n$ certain preimages
of $w^1,\ldots,w^s$ under $\phi$. By Lemma \ref{BP1} each minor of order $\tau$ of $A=(\eta_{\a_i}(w_j))$ is smaller than
$\tau! [D_n(\tilde k)M_{\tilde k}(\psi)]^\tau \tilde r^e,$ which is strictly less than $t^{-\kappa},$ by the choice of $\tilde r$.

\smallskip

Now if $W=\{w^1,\ldots,w^s\}$ is also contained in $\O(t,{\mathbb Z})$, then the entries of $A$ are integers divided by $t$. Therefore the
determinant $\Delta$ of each of the minor of order $\tau$ of $A$ is an integer divided by $t^\kappa.$ Therefore, if $|\Delta|$ is strictly
smaller than $t^{-\kappa}$, it is zero. We conclude that each finite set $W=\{w^1,\ldots,w^s\}$ contained in $\O(t,{\mathbb Z}) \cap \phi(B_j),$
is contained in an algebraic hypersurface of degree less than or equal to $d$.

This fact, together with the estimate above on the number of the covering balls, proves that $\O(t,{\mathbb Z})$ is contained in the union of
at most $C(\psi,d)t^{\e(\psi,d)}$ algebraic hypersurfaces of degree less than or equal to $d$. An accurate evaluation of the constants
(see \cite{Pil3}, Section 4) shows that $\e(\psi,d)\to 0$ as $d\to \infty$. This completes the proof. $\square$

\smallskip

To finally bound the number of points in $\O(t,{\mathbb Z})$ (and not only the number of algebraic hypersurfaces to which these points belong)
an additional tool is used: a kind of ``Bezout theorem'', i.e. an upper bound on the possible number of intersection points of $\O$ with an
algebraic hypersurface of degree $d$ (see \cite{Pil1,Pil2}). In many specific cases such a bound is available. One of the results obtained
in this way in $\cite{Pil3}$, concerns rational points on a compact subanalytic surface $\O \subset {\mathbb R}^n$. Notice that such surface
may contain finite or infinite number of algebraic curves, and it is natural to count rational points on the complement $\O^{trans}$ of these
curves.

\bt\label{Sub.Pil} (\cite{Pil3}, Theorem 1.3) For each $\e>0$ there is a constant $C(\O,\e)$ such that for each $t$ the set
$\O^{trans}(t,{\mathbb Z})$ contains at most $C(\O,\e)t^\e$ points.
\et
We do not attempt to present here more specific results from this exciting field of recent research. They can be found in original papers
\cite{Bom.Pil},\cite{Pil1}-\cite{Pil6},\cite{Pil.Wil} and many other. See also a recent review in \cite{Sca}.

\smallskip

Let us only stress that typically
a proof consists of three parts: a certain smooth parametrization is used, to provide conditions for applicability of Lemma \ref{BP1}.
According to the specific setting, this part may require application of some of smooth parametrization results presented above. In some
other cases, algebraic resolution of singularities or analytic uniformization is used, or the required smooth parametrization is
explicitly constructed for the specific situation considered.

\smallskip

Next, Proposition \ref{BP2} (in one or another form) is applied. It provides a bound on the number of algebraic hypersurfaces of degree
less than or equal to $d$, which contain all the rational points under question.

\smallskip

Finally, an upper bound is produced on the possible number of intersection points in $\O$ with an algebraic hypersurface of degree $d$.
Also here various tools can be applied (compare \cite{Pil1,Pil2}).

\smallskip

A lot of open questions have been discussed in the recent literature. One can expect that progress in the first and the third parts of the
approach may provide answers to some of these questions. Let us complete this section with a brief discussion of one of these questions,
namely, the Wilkie conjecture concerning rational points on exponential-algebraic varieties.

\medskip

\noindent{\bf Conjecture 2} (\cite{Pil.Wil}) {\it Suppose $Y$ is definable in ${\mathbb R}_{exp}$. Then the number of rational points of the
height $T$ in $Y^{trans}$ is at most $c(Y)(\log T)^{C(Y)}$.}

\medskip

A very accurate result for rational on the graph of the Riemann zeta-function was obtained in \cite{Mas}:

\bt\label{Mas}
There is a positive effective absolute constant $C$ such that for any integer $D\geq 3$ the number of rational $z$ with $2<z<3$ of
denominator at most $D$ such that $\zeta(z)$ is rational also of denominator at most $D$ is at most $C({{\log D}\over {\log \log D}})$.
\et
The proof is based on an analytic version of the approach of \cite{Bom.Pil}, and on an accurate estimate of the number of zeroes of a
polynomial $P(z,w)$ of degree $d$ in each of the variables, restricted to the graph of $w=\zeta(z)$. Inside the disk $D_R$ this number
does not exceed $C_1d(d+R \log R)$ (\cite{Mas}, Proposition 1). Notice, that similar results for general analytic functions, with the
quadratic dependence of the number of zeroes on the degree $d$ of $P$ (but, in general, with gaps in the degrees) were obtained in
\cite{Com.Pol}.

\smallskip

Some other specific cases in the direction of Wilkie's conjecture were settled in \cite{Pil6,But,Jon.Tho,Jon.Mil.Tho}. It would be interesting
to compare Theorem \ref{Mas} and Wilkie's conjecture
with Conjecture 1 (by now proved in \cite{Bur.Lia.Yan}) in Section \ref{Dyn.Appl} above, concerning local entropy growth in Analytic
Dynamics. Both the formulations and the methods applied in \cite{Pil6,Jon.Tho,Jon.Mil.Tho} on one side, and in \cite{Bur.Lia.Yan,Yom3} on
the other, look pretty similar (compare, in particular, ``mild'' and ``ultradifferentiable'' functions). This similarity suggests also
the following question: Conjecture 1 was proved in \cite{Bur.Lia.Yan} not only for analytic, but also for quasi-analytic mappings. Can we
hope for similar results in the case of Wilkie's conjecture?

Let us mention a recent paper \cite{Sca1}, devoted to the ``dynamical Mordell-Lang conjecture'' which provides interesting connections 
between Analytic Dynamics, o-minimal structures, and Arithmetics.

\subsection{Polynomial Approximation of Semi-Algebraic Sets}\label{Comp.Geom}

Piecewise-polynomial representation/approximation of geometric objects is one of the main tools in Computational Geometry and Computer
Assisted Design  (see, for example, \cite{Nar,Non.Lin.Com.Geo} and references therein). Subdivision and parametrization are the most
common tools in these domains. In many cases special requirements are imposed, in particular, topological consistence (compare
\cite{Alb.Mou.Tec,Dia.Mou.Rua,Mou.Win}), or an optimal fitting to certain computational requirements
(see \cite{Xu.Mou.Duv.Gal1,Xu.Mou.Duv.Gal2,Xu.Mou.Duv.Gal3}) and references therein.

\smallskip

We believe that control of high order derivatives in many cases may improve performance of subdivision and parametrization methods, to
the extent that justifies additional computational efforts. In particular, this concerns representation of algebraic varieties and
semi-algebraic sets, implicitly given by their equations and inequalities. Conventional accuracy estimates in such approximations include
bounds on the high order derivatives, or on the surface curvature. As an algebraic surface degenerates to a singular one, the curvature
blows up. Consequently, the same happens with the complexity of the approximation: to keep the required accuracy, we need more and more
patches at near-singular (high curvature) areas. This indeed happens in any conventional ``triangulation" algorithm. Moreover, since an
accurate detection of near-singular domains is a complicated problem by itself, mostly certain ``default'' curvature bounds are assumed.
As a result, on one hand, many more than necessary patches are used in smooth areas, while, on the other hand, severe distortions are
produced near singularities.

\smallskip

Smooth parametrization, considered in the present paper, settles (in principle) exactly this problem: a semi-algebraic set $A$ inside the
unit cube, is subdivided into the parts $A_j$, each being covered by a standard ``chart'' with uniform bound on the high order derivatives.
In the case of $C^k$-parametrization the number of the parts $A_j$ depends only on the diagram $D(A)$ and on $k$. The curvature,
which certainly may blow up as non-singular components of $A$ degenerate to singular ones, does not affect at all the complexity of a
$C^k$-parametrization. In \cite{Hav.Yom} some initial results were obtained, concerning application of $C^k$-parametrization in polynomial
approximation of implicit algebraic varieties. Based on a certain setting of singularities analysis, suggested in \cite{Yom6}, we also
suggest in \cite{Hav.Yom} a combined method based on ``of line'' parametrization of certain model surfaces, with their on line fitting
to actual data.

\smallskip

Below we provide explicit and uniform bounds on the complexity of a polynomial $\e$-approximation of $A$ (see Definition \ref{e.appr.comp.def}
below), depending only on the diagram $D(A)$, in two different ways: using either $C^k$-parametrization, or analytic one (only in dimension $2$).
The first bound has a form $K(\sigma, D(A))({1\over \e})^\sigma,$ for each $\sigma > 0$. Another bound has a form $K_2(D(A))(\log({1\over \e})^3.$

\smallskip

Let $A$ be a semi-algebraic set inside the unit cube $I^n \subset {\mathbb R}^n$, and let $D(A)$ be its diagram.

\bd\label{e.appr.def}

A parametric polynomial $(d,\e)$-approximation $\Phi$ of $A$ (or simply a $(d,\e)$-approximation) is a collection of polynomial mappings
$\phi_j: I^{n_j}\to {\mathbb R}^n, \ j=1,\ldots,N,$ (with all the components of $\phi_j$ of degree $d$), satisfying the following
condition:

\smallskip

\noindent There exists a parametrization of $A$, in sense of Definition \ref{rep}, i.e. a subdivision of $A$ into semi-algebraic pieces
$A_j$ together with algebraic mappings $\psi_j: I^{n_j}\to A_j, \ j=1,\ldots,N,$ such that $\psi_j$ are onto and homeomorphic on the
interiors of $I^{n_j}$ and $A_j$, and such that $\max_{x\in I^{n_j}}||\phi_j(x)-\psi_j(x)||\leq \e$ for each $j=1,\ldots,N$. Here the norm
$||.||$ denotes the Euclidean norm in ${\mathbb R}^n$.
\ed
Next we want to compare ``complexities'' of different $(d,\e)$-approximations of $A$. There are many ways to define such complexity
(depending on the application in mind - compare \cite{Yom4} and references therein). We choose one of options suggested in \cite{Yom4},
which naturally appears in many analytic-geometric problems, like the structure of critical sets and values, zero counting, etc:

\bd\label{e.appr.comp.def}
A complexity $C(\Phi)$ of a $(d,\e)$-approximation $\Phi$ of $A$ is equal to the sum $\sum_{j=1}^N d^{n_j}$. An $\e$-complexity $C(\e,A)$
of $A$ is the minimum of $C(\Phi)$ over all $d$ and all the $(d,\e)$-approximations $\Phi$ of $A$.
\ed

\bt\label{Ck.compl.th}
Let $A$ be a semi-algebraic set inside the unit cube $I^n \subset {\mathbb R}^n$, and let $D(A)$ be its diagram. Then for each $\sigma > 0$
there is a constant $K(\sigma,D(A))$ such that for each $\e > 0$ we have

\be\label{Ck.compl.eq}
C(\e,A) \leq K(\sigma,D(A)) ({1\over \e})^{\sigma}.
\ee
\et
\pr
Let $\sigma > 0$ be given. Put $k=[{n\over \sigma}]+1$. Applying $C^k$-parametrization theorem (Theorem \ref{mainck} above) we find
a $C^k$-parametrization of $A$, with the $C^k$-charts $\psi_j: I^{n_j}\to A_j, \ j=1,\ldots,M,$ where $M=M(\sigma,D(A))$, depends only on
$k=[{n\over \sigma}]+1$ and on the diagram $D(A)$ of $A$. Next we subdivide each $I^{n_j}$ into sub-cubes $\tilde I^{n_j}$ of the diameter
$r$. Here $r=c_1 \e^{{1\over k}}$ is chosen in such a way that the remainder term in the Taylor formula of degree $d=k-1$ for each $\psi_j$
is less than $\e$. Defining $\phi_{ji}$ on the sub-cubes $\tilde I_i^{n_j}$ as the corresponding Taylor polynomial mappings, we get
$\max_{x\in I_i^{n_j}}||\phi_j(x)-\psi_{ji}(x)||\leq \e$. Therefore $\Phi,$ consisting of all the polynomials mappings
$\psi_{ji}: \tilde I_i^{n_j}\to {\mathbb R}^n,$ is a $(d,\e)$-approximation of $A$. The number $N$ of the polynomial pieces in $\Phi$ is at
most $M({1\over r})^n={M\over {c^n_1}} ({1\over \e})^{n\over k}\leq  {M\over {c^n_1}}({1\over \e})^{\sigma}.$
Hence $C(\Phi)$ does not exceed $N d^n \leq K(\sigma,D(A)) ({1\over \e})^{\sigma},$ with
$K(\sigma,D(A))={M\over {c^n_1}}d^n={M\over {c^n_1}}[{n\over \sigma}]^n.$ This completes the proof. $\square$.

\bt\label{Anal.compl.th}
Let $A$ be a semi-algebraic set inside the unit cube $I^2 \subset {\mathbb R}^2$, and let $D(A)$ be its diagram. Then for each $\e > 0$ we
have

\be\label{Anal.compl.eq}
C(\e,A) \leq K_2(D(A)) (\log {1\over \e})^3.
\ee
\et
\pr
By the Analytic parametrization theorem (Theorem \ref{anal.rep.th}) we can remove from $I^2$ not more than $N(D(A)$ boxes of size $\e$
and analytically parametrize the remaining part of $A$ with $a$-charts $\psi_j:I^{n_j}\to A_j, \ j=1,\ldots,M$ where
$M=M(D(A))\log {1\over \e}$. Next we choose the degree $d$ of the Taylor polynomials $\phi_j$ of $\psi_j$: by Cauchy formula the polynomial
approximation of degree $d$ of any a-chart $\psi_j,$ provided by a $d$-truncation of its Taylor series, has an accuracy $2^{-d}$ on $I^{n_j}$.
So if we fix $d=[\log{1\over \e}]+1$ we get $\max_{x\in I^{n_j}}||\phi_j(x)-\psi_j(x)||\leq \e$ for each $j=1,\ldots,M.$ For the removed
parts of $A$ (each of a size at most $\e$) we use $C^1$-parametrization, and polynomial approximations of degree zero. The error still
cannot exceed $\e$. For the constructed $(d,\e)$-approximation $\Phi$ of $A$ we have the complexity $C(\Phi)$ equal, according to Definition
\ref{e.appr.comp.def}, to the sum $\sum_{j=1}^M d^{n_j}$. By the estimates above, this sum does not exceed

$$
M(D(A))\log{1\over \e}(\log{1\over \e}+1)^2 + N(D(A)) \leq K_2(D(A)) (\log {1\over \e})^3.
$$
The second term on the left counts the contribution of the removed $\e$-boxes. This completes the proof of Theorem \ref{Anal.compl.th}.
$\square$

\smallskip

\smallskip

Asymptotically, as the allowed error $\e$ tends to zero, the ``analytic'' bound is much better than the $C^k$ one. We conjecture, that it is
optimal. However, for some choices of $\sigma$ and for specific values of $\e$, the $C^k$ bound may turn to be better. An accurate comparison
of the bounds of Theorems \ref{Ck.compl.th} and \ref{Anal.compl.th} below is an important open problem. Notice, that exactly the same
dichotomy appears also in bounding rational points: $C^k$-parametrization leads to $\sigma$-power bound, for each $\sigma > 0$,
while mild parametrization provides logarithmic bound. This problem is related also to the comparison of two approaches to bounding a local
entropy of analytic maps (which both provide logarithmic bound - see Section \ref{Dyn.Appl} above): the $C^k$ with $k\to \infty$ approach of
\cite{Bur.Lia.Yan}, and analytic approach of \cite{Yom3}.

\medskip

\noindent{\bf Conjecture 3} {\it Let $A$ be a semi-algebraic set inside the unit cube $I^n \subset {\mathbb R}^n$, and let $D(A)$ be its
diagram. Then for each $\e > 0$ we have

\be\label{Anal.compl.eq}
C(\e,A) \leq K_n(D(A)) (\log_2{1\over \e})^{n+s(D(A))}.
\ee}
Let us stress that while we consider the results above as really promising, they are still far from any real application. The author is
not aware of any computer implementation of high-order smooth parametrization algorithms.

\subsection{Remez-type inequalities on Algebraic Curves}\label{Remez}

In this section we briefly discuss robustness of polynomial approximation on algebraic curves. This is an important question
in Approximation Theory, and its connection with a sort of analytic parametrization is well known (see \cite{Bar.Ple1}-\cite{Bar.Ple3},
\cite{Pie,Pie1,Pie2,Yom9} and references therein). One of the main tools here is provided by ``Remez-type'' or ``norming'' inequalities, 
which compare maximum of a polynomial on the unit interval $I=[-1,1]$ with its maximum on a given subset $Z\subset I$.

\smallskip

The classical Remez inequality (\cite{Rem}) is as follows:

\bt \label{Rem} Let $P(x)$ be a real polynomial of degree $d$. Then for any measurable $Z\subset [-1,1]$
\be\label{Rem.ineq}
\max_{[-1,1]} \vert P(x) \vert \leq T_d({{4-\mu}\over {\mu}})\max_Z \vert P(x) \vert,
\ee
where $\mu=\mu_1(Z)$ is the Lebesgue measure of $Z$ and $T_d(x)= \cos(d \ \arccos(x))$ is the $d$-th Chebyshev polynomial.
\et

\smallskip

Inequalities of the form (\ref{Rem.ineq}) are known also for sets $Z$ of measure zero, for discrete or finite $Z$
(see \cite{Bru.Yom,Yom8,Yom9} and references therein). Similar inequalities have been studied for restrictions of polynomials
to semi-algebraic (subanalytic) sets (\cite{Bar.Ple1}-\cite{Bar.Ple3}, 
\cite{Bos.Lev.Mil.Tay,Bos.Lev.Mil.Tay,Bos.Bru.Lev,Bru,Com.Pol,Com.Pol1,Pie,Yom9}).
However, in contrast with Theorem \ref{Rem}, already on algebraic curves we cannot hope to get a uniform bound, depending only on the
degree and on the geometry (measure) of $Z$. It is important to stress that this question is directly related to counting zeroes of the
restrictions of polynomials to analytic curves. Specifically, see \cite{Com.Pol,Bru} and a discussion after Theorem \ref{Mas} above.

\smallskip

Let a non-singular real algebraic curve $Y \subset I^2$ be given by the equation $P(x,y)=0$, with $P$ a real polynomial of degree $d$,
and let $Z\subset Y$. We consider restrictions to $Y$ of polynomials $Q(x,y)$ of degree $d_1$.

\bd\label{Remez.def} The Lebesgue (or Remez, or Norming) constant $R_{d_1,Y}(Z)$ is the minimal constant $K$ in the inequality

\be\label{Remez.eq}
\max_{(x,y)\in Y} |Q(x,y)| \leq K \max_{(x,y)\in Z} |Q(x,y)|,
\ee
valid for polynomials $Q(x,y)$ of degree $d_1$.
\ed

Let us consider the following example. Our curve $Y$ is the part of the hyperbola $Y_\e=\{xy=\e^2\}$ inside the square
$I^2_1=[0,1]\times [0,1] \subset {\mathbb R}^2.$ Put $Z_\e$ be the half-branch of $Y$ with $\e \leq x \leq 1$, and consider a polynomial
$Q(x,y)=y$ on $Y_\e$. We have $\max_{Z_\e} |Q(x,y)|=\e$, while $\max_{Y_\e} |Q(x,y)|=1$. Therefore the Remez constant
$R_{1,Y_\e}(Z_\e)$ for $Z_\e$ on $Y_\e$ is at least $1\over \e$. But the length of $Z_\e$ is exactly one half of the length of $Y_\e$. Hence
the Remez type inequality on $Y_\e$ cannot be uniform (in contrast with the classical case): it depends not only on the degrees of the
polynomials $P$ and $Q$ involved, but also on the value of the coefficient $\e$.

\smallskip

Analytic parametrization can be used in Remez type inequalities as follows: assume that the images of the the a-charts provide a ``uniformly
overlapping'' covering of $Y$, and that on these images a uniform Remez type inequality, with the Remez constant, say, equal to $2$ is satisfied
(compare \cite{Yom9}, Section 5. We call such parametrization a ``Remez'' one). Both these conditions can be satisfied with a relatively small
modification of the constructions used in Section \ref{An2} above. Then in order to get a global Remez type inequality on $Y$ we can extend it 
along the chains of the a-charts (compare \cite{Yom9}, Theorem 5.2.) The expected upper bound for the Remez constant on $Y$ will be $2^N$, where 
$N$ is the number of the charts.

\smallskip

In the example of the hyperbola $Y_\e$ we know that the number of covering charts is of order $N=\log {1\over \e}$ (\cite{Yom7}), so the
expected bound $2^N={1\over \e}$ is sharp.

\smallskip

The following theorem extends this observation from the hyperbola $Y_\e$ to any regular real algebraic curve $Y \subset I^2_1$:

\bt\label{Remez.curves.th}
Let a non-singular real algebraic curve $Y \subset I^2_1$ be given by the equation $P(x,y)=0$, with $P$ a real polynomial of degree $d,$ with
the norm $||P||=\max_{I^2_1}|P|$ equal to $1$. Put $\rho = \min_{(x,y)\in Y} ||grad P(x,y)|| >0$. Then there exists a Remez
analytic parametrization of $Y$ with at most $N=C(d)\log {1\over \rho}$ a-charts.
\et
\pr
We provide a sketch of the proof. Using the assumption on the gradient of $P$ and applying a ``quantitative Implicit
Function theorem'' (see, for example, \cite{Yom5}) we can find for each point $(x,y)$ on $Y$ a coordinate chart centered at this point, which
is, in fact, an a-chart, and which covers a $2\delta = c_1(d)\rho$-neighborhood of $(x,y)$ on $Y$. Next, we apply to the curve $Y \subset I^2_1$
an analytic parametrization Theorem \ref{anal.rep.th}, and cover $Y$ except its intersection with at most $N_1(d)$ \ $\delta$-cubes $Q_j$ in
$I^2_1$. The number of a-charts as at most $N_2(d)\log {1\over \delta}=N_2(d)\log {1\over {c_1(d)\rho}}.$ Finally, we cover $Y$ inside each
$Q_j$ with exactly one coordinate chart, as constructed above. $\square$.

\smallskip

The connection between the analytic parametrization of $Y$ and the bound on the Remez constant $R_{d_1,Y}(Z)$ for subsets $Z$ of $Y$ was
outlined above: we can extend the Remez bound along the chains of a-charts. Accordingly, as the coefficients of the polynomial $P$ defining
$Y$ vary, we can expect the Remez constant to behave as $2^N=({1\over \rho})^{C(d)}$, where $\rho$ is the minimum of the norm of the gradient
of $P$ on $Y$. We plan to present detailed results in this direction separately.

\smallskip

An interplay between real and complex geometry is an essential part of the real analytic parametrization, and real Remez-type inequalities.
Interestingly, there are similar effects in Dynamics - compare \cite{Mon}. Are there direct connections?

Let us conclude this section with an informal conjecture that results similar to Theorem \ref{Remez.curves.th} are valid also in higher
dimensions.

\subsection{Signal Sampling and Motion Planning}

In Algebraic signal sampling some difficult and practically very important problems lead to a necessity to process in a robust way some highly
degenerated algebraic singularities. In particular, this happens in Fourier reconstruction of ``spike trains'', i.e. of linear combinations
of $\delta$-functions $F(x)=\sum_{j=1}^d a_j\delta(x-x_j$. Here the amplitudes $a_j$ and the positions $x_j$ are unknown, and have to be found
from the Fourier samples of $F$. This problem becomes notoriously difficult, as the nodes ``almost collide'', especially in the presence of a
realistic noise. The Algebraic Geometry of the nodes collision singularity turns out to be very complicated - compare \cite{Bat.Yom1} and 
references therein. We believe that application of smooth parametrization to the corresponding near-singular varieties may significantly
improve the reconstruction accuracy and resolution. 

Motion Planning in Robotics presents another plausible application of smooth parametrizations. Also here Algebraic Geometry provides an adequate
description of the problem, and near-singularities cannot be avoided (compare \cite{Eli.Yom} and references therein). A natural requirements is 
that the higher derivatives of the motion (acceleration, jerk) remain bounded. Once more, smooth parametrization may provide a required control.

\medskip



\bibliographystyle{amsplain}

\end{document}